\documentclass[showkeys,showpacs, aps,amssymb,keywords,floatfix,prd,amsmath,preprintnumbers]{revtex4}
\RequirePackage{float}
\usepackage{epstopdf}
\usepackage{hyperref}
\usepackage{capt-of}
\usepackage{graphicx}  % Include figure files
\usepackage{dcolumn}   % Align table columns on decimal point
\usepackage{bm}% bold math
\usepackage{booktabs}
\usepackage{nicefrac} 
\usepackage{mathtools}
\usepackage{resizegather}
\begin{document}
	
%\input epsf.tex
%%%%%%%%%%%%
%%%%%%%%%%%
\title{Inflation in mimetic $f(R,T)$ gravity}

\author{Snehasish Bhattacharjee\footnote{Email: snehasish.bhattacharjee.666@gmail.com}
}

\affiliation{Department of Physics, Indian Institute of Technology, Hyderabad 502285, India}
\date{\today}

\begin{abstract}
In this paper, we employ mimetic $f(R,T)$ gravity coupled with Lagrange multiplier and mimetic potential to yield viable inflationary cosmological solutions consistent with latest Planck and BICEP2/Keck Array data. We present here three viable inflationary solutions of the Hubble parameter ($H$) represented by $H(N)=\left(A \exp \beta  N+B \alpha ^N\right)^{\gamma }$, $H(N)=\left(A \alpha ^N+B \log N\right)^{\gamma }$, and $H(N)=\left(A e^{\beta  N}+B \log N\right)^{\gamma }$, where $A$, $\beta$, $B$, $\alpha$, $\gamma$ are free parameters, and $N$ represents the number of e-foldings. We carry out the analysis with the simplest minimal $f(R,T)$ function of the form $f(R,T)= R + \chi T$, where $\chi$ is the model parameter. We report that for the chosen $f(R,T)$ gravity model, viable cosmologies are obtained compatible with observations by conveniently setting the Lagrange multiplier and the mimetic potential.
\end{abstract}



\maketitle

\section{Introduction}
Inflation has been remarkably successful in sufficing some of the most important cosmological enigmas such as the flatness problem, horizon problem and the fine-tuning problems \cite{in1,in2,in3}. Furthermore, it has also been argued that if the pre-inflationary density fluctuations are allowed to grow super-exponentially over a short period ($ \sim 10^{-33}$ sec), it provides a clear and coherent description of the formation of the large scale structures and the CMB anisotropies \cite{in4,in5,in6,in7}. \\
In Einstein's gravity, a scalar field called "inflaton" is predicted to be responsible for carrying out an inflation \cite{in1}. Numerous potentials have been reported to give rise to an inflation consistent with observations \cite{in8,in9,in10,in11,in12}. \\
Modified theories of gravity have been very productive in delineating cosmological problems such as the flat rotation curves of spiral galaxies or the late-time cosmic acceleration without requiring dark matter and dark energy \cite{bp2,bp3}. Modified theories of gravity give rise to interactions that are scale-dependent but conserve the results of GR at the Solar System scales \cite{bp2}. Thus, Einstein's gravity can be considered as being a specific case of a more diverse set of gravitational theories \cite{bp}.\\
In addition to scalar fields, viable inflationary scanarios have been reported in various modified theories of gravity. Geometric models of inflation have been investigated in $f(R)$ gravity \cite{in13to16}, teleparallel gravity \cite{in17}, Einstein Gauss bonnet gravity \cite{in18}, $f(R,T)$ gravity \cite{bhatta,frt2}, and also in braneworld models \cite{in19}.\\
Mimetic gravity is a well-organized Weyl-symmetric augmentation of Einstein's gravity connected through a singular disformal transformation capable of presenting a unified description of the evolution of the Universe from the primordial inflation to the late-time acceleration \cite{mimetic}. Mimetic gravity theories have been employed to investigate inflation \cite{framework,fg,fr2,noj,sd1,sd2,sd3}, black holes solutions \cite{black}, late-time acceleration \cite{mimetic,ltm,noj}, bouncing cosmology \cite{noj}, cosmological perturbations \cite{sd4}, rotation curves of galaxies and wormholes \cite{rot}.\\ 
The purpose of the present article is to investigate inflationary cosmological solutions consistent with latest Planck and BICEP2/Keck Array data in the framework of mimetic $f(R,T)$ gravity with Lagrange multiplier and mimetic potential. $f(R,T)$ gravity was introduced in \cite{in20} and has been extremely successful in the understanding of the flat rotation curves \cite{in22}, late-time acceleration \cite{in20}, baryogenesis \cite{in40}, bouncing cosmology \cite{in37,in38}, evolution of density perturbations \cite{in42}, redshift drift \cite{in41}, Big bang nucleosynthesis \cite{in44}, temporally varying physical constants \cite{in43}, inflation \cite{bhatta,frt2}, and viscous cosmology \cite{viscous}. \\
The manuscript is organized as follows: In Section \ref{sec2}, we provide an overview of mimetic $f(R,T)$ gravity. In Section \ref{sec3}, we present the inflationary models and discuss viable cosmologies within the framework of mimetic $f(R,T)$ gravity with Lagrange multiplier and mimetic potential. Finally, in Section \ref{sec4} we present the results and conclusions.  

\section{Mimetic $f(R,T)$ gravity}\label{sec2}
In this section we shall present an overview of mimetic $f(R,T)$ gravity with Lagrange multiplier and mimetic potential. Our aim to investigate inflation in mimetic $f(R,T)$ gravity stems from the fact that a broader class of solutions is possible by parametrizing the metric using novel degrees of freedom \cite{mmfrt,mm22,mm33,mm34,mm35,mm36,mm37,mm38,mm39}. We shall therefore express the metric $g_{ij}$ as a function of an auxiliary metric $\widetilde{g}_{ij}$ along with an auxiliary scalar field $\phi$, as follows,
\begin{equation}\label{eq1}
g_{ij}=-\widetilde{g}_{ij} \partial_{\alpha} \partial_{\beta} \phi \widetilde{g}^{\alpha\beta}.
\end{equation}
From Eq. \ref{eq1}, we obtain the identity
\begin{equation}\label{eq2}
-(\widetilde{g}_{ij}, \phi) \partial_{i} \phi \partial_{j}\phi g^{ij}=1.
\end{equation}
Since we are interested in investigating inflationary scenarios in mimetic $f(R,T)$ gravity, we shall therefore assume a flat FRW spacetime with $'-','+','+','+'$ metric signature. Furthermore, we assume a lagrange multiplier $\lambda(\phi)$ along with a mimetic potential $V(\phi)$. The action therefore reads \cite{mmfrt}
\begin{equation}\label{eq3}
S = \int \sqrt{-g}dx^{4}\left(\mathcal{L}_{m} - V(\phi) + \lambda(g_{ij} \partial_{i}\phi \partial_{j}\phi +1) + f(R(g^{ij},T))\right), 
\end{equation}
where $R$ and $T$ denote respectively the Ricci scalar and trace of the energy-momentum tensor, and $\mathcal{L}_{m}$ represent the matter Lagrangian. \\
Varying the action (Eq. \ref{eq3}) with respect to the metric $g_{ij}$ leads to the following field equation 
\begin{equation}\label{eq4}
\frac{1}{2}g_{ij}\left(\lambda (g^{\alpha\beta} \partial_{\alpha}\phi \partial_{\beta}\phi +1) - V(\phi) \right) +\frac{1}{2}g_{ij}f(R,T) + \bigtriangledown_{i}\bigtriangledown_{j}f_{R} - R_{ij}f_{R}-\lambda \partial_{i}\phi \partial_{j}\phi + \frac{1}{2}T_{ij} - f_{T}(T_{ij}+\Theta_{ij}) - g_{ij}\square f_{R} =0,
\end{equation}
where 
\begin{equation}
\Theta_{ij} \equiv g^{\alpha\beta}\frac{\delta T_{\alpha\beta}}{\delta g^{ij}},
\end{equation}
$\square =\bigtriangledown^{i}\bigtriangledown_{i} $ denote the d'Alembertian operator, $\bigtriangledown_{i} $ denote the covariant derivative with respect to the metric and $f_{R}$ and $f_{T}$ denote the partial derivatives of the $f(R,T)$ gravity models with respect to $R$ and $T$ respectively.\\
We shall assume the cosmos to be filled with a perfect fluid and therefore the energy-momentum tensor can be written as
\begin{equation}
T_{ij}=p_{m}g_{ij}+(p_{m}+\rho_{m})v_{i}v_{j},
\end{equation} 
where $p_{m}$, and $\rho_{m}$ represent pressure and energy density of the matter sources respectively and $v_{i}$ denote the four-velocity. We assume the matter Lagrangian $\mathcal{L}_{m} = -p_{m}$. Therefore, we obtain 
\begin{equation}\label{eq5}
\frac{1}{2}g_{ij}\left(\lambda (g^{\alpha\beta} \partial_{\alpha}\phi \partial_{\beta}\phi +1) - V(\phi) \right) +\frac{1}{2}g_{ij}f(R,T) + \bigtriangledown_{i}\bigtriangledown_{j}f_{R} - R_{ij}f_{R}-\lambda \partial_{i}\phi \partial_{j}\phi + \frac{1}{2}T_{ij} - f_{T}(T_{ij}+p_{m}g_{ij}) - g_{ij}\square f_{R} =0.
\end{equation}
One may note that upon varying the action (\ref{eq4}) with respect to the auxiliary scalar field $\phi$, we arrive at
\begin{equation}
- V^{'}(\phi)-2 \bigtriangledown^{i} (\lambda \partial_{i} \phi) =0,
\end{equation}
where prime denote the derivative with respect to $\phi$. Next, if we vary the action with respect to the Lagrange multiplier $\lambda$, we end up with 
\begin{equation}
g^{ij}\partial_{i}\phi \partial_{j}\phi = -1,
\end{equation}
clearly illustrating the fact that the scalar field is not a degree of freedom \cite{noj}. \\
For the FRW background along with an added presumption that the scalar field $\phi$ depends only on time $t$, the Friedmann equations are given as \cite{mmfrt}
\begin{equation}\label{eq6}
6 (\dot{H} +H^{2})f_{R} - f(R,T) - 6 H\frac{d f_{R}}{dt} - \lambda (\dot{\phi}^{2} +1) - 2 f_{T} p_{m} + V(\phi) + \rho_{m} (2 f_{T} +1)=0,
\end{equation} 
\begin{equation}
f(R,T) - 2(\dot{H} +3H^{2})f_{R}+ p_{m} (4 f_{T} +1 + 4 H\frac{d f_{R}}{dt}) - \lambda (\dot{\phi}^{2} +1) + 2 \frac{d^{2}f_{R}}{d t^{2}}  - V(\phi)  
=0,
\end{equation}
\begin{equation}
6 H \lambda \dot{\phi} - V^{'}(\phi) + 2\frac{d(\lambda \dot{\phi})}{dt}=0,
\end{equation}
and
\begin{equation}
\dot{\phi}^{2}-1=0,
\end{equation}
where overhead dot represents time derivative. Since we have assumed $\phi = \phi(t)$, therefore, the equation \ref{eq6} can be written in a more simplified way as follows
\begin{equation}
-2 (\dot{H} +3H^{2})f_{R} + f(R,T) + 4 H\frac{d f_{R}}{dt}+2 \frac{d^{2}f_{R}}{d t^{2}} - V(t) + p_{m} (4 f_{T} +1)=0.
\end{equation}
Therefore, the mimetic potential $V(t)$ in $f(R,T)$ gravity reads \cite{mmfrt}
\begin{equation}\label{eq7}
V(t) =  f(R,T) + 4 H\frac{d f_{R}}{dt}+2 \frac{d^{2}f_{R}}{d t^{2}} + p_{m} (4 f_{T} +1)-2 (\dot{H} +3H^{2})f_{R}.
\end{equation}
By choosing suitable $f(R,T)$ gravity models, one can reconstruct the expressions for the mimetic potential from Eq. \ref{eq7}. Upon solving for the mimetic potential, one can then solve equation \ref{eq6} with respect to Lagrange multiplier $\lambda(t)$ \cite{mmfrt}
\begin{equation}\label{eq8}
\lambda (t) = 3 (\dot{H} +H^{2})f_{R} - \frac{1}{2} f (R,T) -3 H\frac{d f_{R}}{dt} + \rho_{m}(f_{T} + \frac{1}{2}) + \frac{1}{2}V(t) - f_{T} p_{m}.
\end{equation}
\section{Inflationary models in mimetic $f(R,T)$ gravity}\label{sec3}
We shall now turn our attention on illustrating the viability of three distinct inflationary models within the framework of mimetic $f(R,T)$ gravity compatible with the latest Planck and BICEP2/Keck Array results \cite{frame18}. To keep things simple, we shall therefore employ the simplest minimal $f(R,T)$ gravity model of the form $f(R,T)=R+\chi T$, where $\chi$ is a constant. \\
The slow-roll parameters are more conveniently expressed in terms of the e-folding number $N$, rather than time $t$. Therefore, we use the following transformations, 
\begin{equation}\label{eq9}
\frac{d}{dt}=H(N)\frac{d}{dN}, \frac{d^{2}}{dt^{2}}=H(N)\frac{dH}{dN}\frac{d}{dN}+H^{2}(N)\frac{d^{2}}{dN^{2}}.
\end{equation}
Now, following the prescriptions of \cite{frame21}, slow-roll parameters can be written as
\begin{equation}\label{slow1}
\epsilon=-\frac{H(N)}{4H^{'}(N)}\left[\frac{6\frac{H^{'}(N)}{H(N)}+\frac{H^{''}(N)}{H(N)}+\left( \frac{H^{'}(N)}{H(N)}\right)  ^{2}}{\frac{H^{'}(N)}{H(N)}+3} \right] ^{2},
\end{equation}
and
\begin{equation}\label{slow2}
\eta = -\left[\frac{\frac{1}{2}\left( \frac{H^{'}(N)}{H(N)}\right) ^{2}+9\frac{H^{'}(N)}{H(N)}+3\frac{H^{''}(N)}{H(N)}+3\frac{H^{''}(N)}{H^{'}(N)}-\frac{1}{2}\left(\frac{H^{''}(N)}{H^{'}(N)} \right) ^{2}+\frac{H^{'''}(N)}{H^{'}(N)} }{2\left( \frac{H^{'}(N)}{H(N)}+3\right) } \right], 
\end{equation}
where primes denotes derivative with respect to the number of e-folding $N$. Next, the inflation related observables can be written as
\begin{equation}\label{inf}
n_{s}\simeq 1- 6\epsilon + 2\eta, \hspace{0.25in}\text{and}\hspace{0.25in} r\simeq 16 \epsilon,
\end{equation}
where $n_{s}$, and $r$ denote respectively the spectral index of primordial curvature perturbations and the scalar-to-tensor ratio. We aim to find analytical expressions for the mimetic potential $\phi$ and Lagrange multiplier $\lambda$ in mimetic $f(R,T)$ gravity for various inflationary models constrained from the latest Planck and BICEP2/Keck Array data. Recent Planck data placed tight constraints on the inflation observables as \cite{frame18} 
\begin{equation}\label{obs1}
n_{s}= 0.9644 \pm 0.0049, \hspace{0.25in}\text{and}\hspace{0.25in} r<0.10,
\end{equation}
while the BICEP2/Keck Array imposed an even tighter constraint on the scalar-to-tensor ratio as 
\begin{equation}\label{obs2}
r<0.07, \hspace{0.25in}\text{at 95 \% confidence level}.
\end{equation}
In the following subsections, we shall present viable cosmologies with three different inflationary models in tune with the observations (\ref{obs1} and \ref{obs2}). 
\subsection{Inflationary Model I}
One of the simplest inflationary models can be expressed as the following,
\begin{equation}\label{1st}
H(N)=\left(A \exp \beta  N+B \alpha ^N\right)^{\gamma },
\end{equation}
where $A$, $\beta$, $B$, $\alpha$ and $\gamma$ are free parameters. Combining Eq. \ref{1st} with \ref{slow1} and \ref{slow2}, the slow-roll parameters for the model read
\begin{equation}\label{1ste}
\epsilon=-\frac{\gamma  \left(2 A^2 \beta  (\beta  \gamma +3) e^{2 \beta  N}+A B \alpha ^N e^{\beta  N} (\log (\alpha ) (\log (\alpha )+\beta  (4 \gamma -2)+6)+\beta  (\beta +6))+2 B^2 \log (\alpha ) \alpha ^{2 N} (\gamma  \log (\alpha )+3)\right)^2}{4 \left(A e^{\beta  N}+B \alpha ^N\right) \left(A \beta  e^{\beta  N}+B \log (\alpha ) \alpha ^N\right) \left(A (\beta  \gamma +3) e^{\beta  N}+B \alpha ^N (\gamma  \log (\alpha )+3)\right)^2},
\end{equation} 
and
\begin{equation}
\eta=\frac{\splitfrac{A^2 \beta ^3 e^{2 \beta  N} \left(A B \alpha ^N e^{\beta  N} (\beta  (0.5\, -2.5 \gamma )-6. \gamma -1.5)+A^2 \gamma  (-2. \beta  \gamma -6.) e^{2 \beta  N}+(-0.25 \beta -1.5) B^2 \alpha ^{2 N}\right)   +B \log (\alpha ) \alpha ^N }{ \times \left(\splitfrac{\splitfrac{ \splitfrac{ -8. A^3 \beta  e^{3 \beta  N} (\log (\alpha ) (0.0625 \log (\alpha )+\beta  (0.3125 \gamma -0.1875)+0.1875)}{+\beta  (\beta  (\gamma  (1. \gamma -0.625)+0.1875)+2.25 \gamma -0.375))}}{   -5. A^2 B \alpha ^N e^{2 \beta  N} \left( \splitfrac{ \log (\alpha ) (\log (\alpha ) (0.05 \log (\alpha )+\beta  (1. \gamma -0.2)+0.3)}{+ \beta  (\beta  (\gamma  (2.4 \gamma -2.)+0.3)+3.6 \gamma -0.3))+\beta ^2 (\beta  (1. \gamma -0.2)+3.6 \gamma -0.3)}\right)}}{  -0.5 A B^2 \alpha ^{2 N} e^{\beta  N} \left( \splitfrac{ \log (\alpha ) (\log (\alpha ) ((5. \gamma -1.) \log (\alpha )+\beta  (\gamma  (16. \gamma -10.)+3.)}{+12. \gamma +3.)+\beta  (\beta  (5. \gamma -3.)+36. \gamma -6.))+(1. \beta +3.) \beta ^2} \right)} + \splitfrac{ B^3 \gamma  \log ^2(\alpha ) \alpha ^{3 N}}{ (-2. \gamma  \log (\alpha )-6.)}\right)}}{\splitfrac{\left(A e^{\beta  N}+B \alpha ^N\right) \left(A \beta  e^{\beta  N}+B \log (\alpha ) \alpha ^N\right)^2}{ \left(A (\beta  \gamma +3.) e^{\beta  N}+B \gamma  \log (\alpha ) \alpha ^N+3. B \alpha ^N\right)}}.
\end{equation}
Now, by using Eq. \ref{inf}, the inflation observables for the model can be written as 
\begin{gather}
r= 16 \left[-\frac{\gamma  \left(2 A^2 \beta  (\beta  \gamma +3) e^{2 \beta  N}+A B \alpha ^N e^{\beta  N} (\log (\alpha ) (\log (\alpha )+\beta  (4 \gamma -2)+6)+\beta  (\beta +6))+2 B^2 \log (\alpha ) \alpha ^{2 N} (\gamma  \log (\alpha )+3)\right)^2}{4 \left(A e^{\beta  N}+B \alpha ^N\right) \left(A \beta  e^{\beta  N}+B \log (\alpha ) \alpha ^N\right) \left(A (\beta  \gamma +3) e^{\beta  N}+B \alpha ^N (\gamma  \log (\alpha )+3)\right)^2} \right], 
\end{gather}
and 
\begin{gather}
n_{s}= \notag 1- 6\left[-\frac{\gamma  \left(\splitfrac{2 A^2 \beta  (\beta  \gamma +3) e^{2 \beta  N}+A B \alpha ^N e^{\beta  N} (\log (\alpha ) (\log (\alpha )+\beta  (4 \gamma -2)+6)}{+\beta  (\beta +6))+2 B^2 \log (\alpha ) \alpha ^{2 N} (\gamma  \log (\alpha )+3)}\right)^2}{4 \left(A e^{\beta  N}+B \alpha ^N\right) \left(A \beta  e^{\beta  N}+B \log (\alpha ) \alpha ^N\right) \left(A (\beta  \gamma +3) e^{\beta  N}+B \alpha ^N (\gamma  \log (\alpha )+3)\right)^2} \right]  \\ + 2\left[\frac{\splitfrac{A^2 \beta ^3 e^{2 \beta  N} \left(A B \alpha ^N e^{\beta  N} (\beta  (0.5\, -2.5 \gamma )-6. \gamma -1.5)+A^2 \gamma  (-2. \beta  \gamma -6.) e^{2 \beta  N}+(-0.25 \beta -1.5) B^2 \alpha ^{2 N}\right)   +B \log (\alpha ) \alpha ^N }{ \times \left(\splitfrac{\splitfrac{ \splitfrac{ -8. A^3 \beta  e^{3 \beta  N} (\log (\alpha ) (0.0625 \log (\alpha )+\beta  (0.3125 \gamma -0.1875)+0.1875)}{+\beta  (\beta  (\gamma  (1. \gamma -0.625)+0.1875)+2.25 \gamma -0.375))}}{   -5. A^2 B \alpha ^N e^{2 \beta  N} \left( \splitfrac{ \log (\alpha ) (\log (\alpha ) (0.05 \log (\alpha )+\beta  (1. \gamma -0.2)+0.3)}{+ \beta  (\beta  (\gamma  (2.4 \gamma -2.)+0.3)+3.6 \gamma -0.3))+\beta ^2 (\beta  (1. \gamma -0.2)+3.6 \gamma -0.3)}\right)}}{  -0.5 A B^2 \alpha ^{2 N} e^{\beta  N} \left( \splitfrac{ \log (\alpha ) (\log (\alpha ) ((5. \gamma -1.) \log (\alpha )+\beta  (\gamma  (16. \gamma -10.)+3.)}{+12. \gamma +3.)+\beta  (\beta  (5. \gamma -3.)+36. \gamma -6.))+(1. \beta +3.) \beta ^2} \right)} + \splitfrac{ B^3 \gamma  \log ^2(\alpha ) \alpha ^{3 N}}{ (-2. \gamma  \log (\alpha )-6.)}\right)}}{\splitfrac{\left(A e^{\beta  N}+B \alpha ^N\right) \left(A \beta  e^{\beta  N}+B \log (\alpha ) \alpha ^N\right)^2}{ \left(A (\beta  \gamma +3.) e^{\beta  N}+B \gamma  \log (\alpha ) \alpha ^N+3. B \alpha ^N\right)}} \right]. 
\end{gather}
It is now indispensable to ascertain distinct values for the free parameters which could generate viable inflationary scenarios compatible with observations (\ref{obs1} and \ref{obs2}). Bearing that in mind, we proceed with the following combination,
\begin{equation}
\beta=0.035, \hspace*{0.25in} \alpha=1, \hspace*{0.25in} \gamma=2, \hspace*{0.25in} A= -0.005, \hspace*{0.25in} \text{and} \hspace*{0.25in} B=10,
\end{equation}
which yield the following values of the inflation related observables, 
\begin{equation}
r=0.00464576, \hspace*{0.25in} \text{and} \hspace*{0.25in} n_{S}=0.964068.
\end{equation}
It can be noted that the predicted values of the inflation-related observables are well within the strict observational constraints, and therefore ensure the viability of this inflationary model. We are now interested in finding the expressions for the mimetic potential and the Lagrange multiplier responsible for producing the inflationary model \ref{1st}. Employing \ref{1st} in \ref{eq7} and in \ref{eq8}, the expressions for the corresponding mimetic potential and the Lagrange multiplier in $f(R,T)$ gravity read respectively as, 
\begin{equation}
V = 6 \left(2 \left(A e^{\beta  N}+B \alpha ^N\right)^{2 \gamma }+\gamma  \left(A e^{\beta  N}+B \alpha ^N\right)^{2 \gamma -1} \left(A \beta  e^{\beta  N}+B \log (\alpha ) \alpha ^N\right)\right)\\-2 \left(3 \left(A e^{\beta  N}+B \alpha ^N\right)^{2 \gamma }+\gamma  \left(A e^{\beta  N}+B \alpha ^N\right)^{2 \gamma -1} \left(A \beta  e^{\beta  N}+B \log (\alpha ) \alpha ^N\right)\right)+4 \rho  \chi -\rho  (4 \chi +1).
\end{equation}
and
\begin{equation}
\lambda = -0.5 \left(6 \left(2 \left(A e^{\beta  N}+B \alpha ^N\right)^{2 \gamma }+\gamma  \left(A e^{\beta  N}+B \alpha ^N\right)^{2 \gamma -1} \left(A \beta  e^{\beta  N}+B \log (\alpha ) \alpha ^N\right)\right)+4 \rho  \chi \right)\\\notag +0.5 \left( 6 \left(2  \left( A e^{\beta  N}+B \alpha ^N\right)^{2 \gamma }+\gamma  \left(A e^{\beta  N}+B \alpha ^N\right)  ^{2 \gamma -1} \left(A \beta  e^{\beta  N}+B \log (\alpha ) \alpha ^N\right)\right)-2 \left(3 \left(A e^{\beta  N}+B \alpha ^N\right)^{2 \gamma }+\gamma  \left(A e^{\beta  N}+B \alpha ^N\right)^{2 \gamma -1} \left(A \beta  e^{\beta  N}+B \log (\alpha ) \alpha ^N \right)\right)+4 \rho  \chi -\rho  (4 \chi +1) \right)\\ \notag +3 \left(\left(A e^{\beta  N}+B \alpha ^N\right)^{2 \gamma }+\gamma  \left(A e^{\beta  N}+B \alpha ^N\right)^{2 \gamma -1} \left(A \beta  e^{\beta  N}+B \log (\alpha ) \alpha ^N\right)\right)+\rho  \chi +\rho  (\chi +0.5).
\end{equation}
\subsection{Inflationary Model II}
Another viable inflationary model which can be employed to demonstrate the usefulness of mimetic $f(R,T)$ grvity and compatible with the both Planck and BICEP2/Keck Array data can be written as  
\begin{equation}\label{2nd}
H(N)=\left(A \alpha ^N+B \log N\right)^{\gamma },
\end{equation}
where $A$, $\alpha$, $B$ and $\gamma$ are the fitting parameters. Employing Eq. \ref{2nd} with \ref{slow1} and \ref{slow2}, the slow-roll parameters for this model reads 
\begin{equation}
\epsilon=-\frac{\gamma  \left(2 A^2 N^2 \log (\alpha ) \alpha ^{2 N} (\gamma  \log (\alpha )+3)+A B \alpha ^N (N \log (\alpha ) (4 \gamma +N (\log (\alpha )+6) \log (N)-2)+6 N-1)+B^2 (2 \gamma +(6 N-1) \log (N)-1)\right)^2}{4 N \left(A \alpha ^N+B \log (N)\right) \left(A N \log (\alpha ) \alpha ^N+B\right) \left(A N \alpha ^N (\gamma  \log (\alpha )+3)+B \gamma +3 B N \log (N)\right)^2},
\end{equation}
and
\begin{equation}
\eta =\frac{\splitfrac{ \splitfrac{ A^3 B N \log (\alpha ) \alpha ^{3 N}}{ \left( \splitfrac{ \splitfrac{ N \log (\alpha ) (N \log (\alpha ) (\gamma  (5.\, -8. \gamma )}{+N \log (N) ((0.5\, -2.5 \gamma )}}{\splitfrac{ \log (\alpha )-6. \gamma -1.5)-1.5) +N (3.\, -18. \gamma )}{+2.5 \gamma -1.5)+1.5 N-1.}}\right) }}{ \splitfrac{ +A^2 B^2 \alpha ^{2 N}}{ \left(N \log (\alpha )  \left(\splitfrac{\splitfrac{ N (\gamma  (12.5\, -12. \gamma )-3.5) \log (\alpha )}{+\log (N) (N \log (\alpha ) (N (1.\, -5. \gamma ) \log (\alpha )}}{\splitfrac{ \splitfrac{ +2.5 \gamma -18. \gamma  N+1.5 N-2.)+3. N-2.)}{+5. \gamma +N^3 (-0.25 \log (\alpha )-1.5)}}{ \log ^2(\alpha ) \log ^2(N)-18. \gamma  N+4.5 N-2.}}\right)  +1.5 N-0.75\right)}}\splitfrac{ +A B^3 \alpha ^N }{\left(\splitfrac{\splitfrac { \splitfrac { \log (N) (N \log (\alpha ) (N (1.\, -2.5 \gamma )}{\times \log (\alpha )+5. \gamma -18. \gamma  N}}{\splitfrac{ +4.5 N-2.)+3. N-1.5)+N (\gamma }{ (10.\, -8. \gamma )-3.) \log (\alpha )}}}{\splitfrac{+2.5 \gamma +N \log (\alpha ) \log ^2(N) (N \log (\alpha ) (-0.5 N \log (\alpha )-1.5 N-0.5)}{+1.5 N-1.)-6. \gamma  N+1.5 N-1.}} \right)}+\splitfrac{\splitfrac{ B^4 (\gamma  (2.5\, -2. \gamma )+\log (N) }{\times(N (1.5\, -6. \gamma )+2.5 \gamma }}{\splitfrac{+(1.5 N-0.75) \log (N)-1.)-0.75)+A^4 \gamma  N^4}{ \log ^3(\alpha ) \alpha ^{4 N} (-2. \gamma  \log (\alpha )-6.)}}}{\splitfrac{ N \left(A \alpha ^N+B \log (N)\right) \left(A N \log (\alpha ) \alpha ^N+B\right)^2}{\times \left(A \gamma  N \log (\alpha ) \alpha ^N+3. A N \alpha ^N+B \gamma +3. B N \log (N)\right)}}.
\end{equation} 
Therefore the inflationary observables (Eq. \ref{inf}) for this model are
\begin{gather}
r= 16 \left[ -\frac{\gamma  \left(2 A^2 N^2 \log (\alpha ) \alpha ^{2 N} (\gamma  \log (\alpha )+3)+A B \alpha ^N (N \log (\alpha ) (4 \gamma +N (\log (\alpha )+6) \log (N)-2)+6 N-1)+B^2 (2 \gamma +(6 N-1) \log (N)-1)\right)^2}{4 N \left(A \alpha ^N+B \log (N)\right) \left(A N \log (\alpha ) \alpha ^N+B\right) \left(A N \alpha ^N (\gamma  \log (\alpha )+3)+B \gamma +3 B N \log (N)\right)^2}\right],
\end{gather}
and
\begin{gather}
n_{s}= \notag 1- 6\left[-\frac{\gamma  \left(2 A^2 N^2 \log (\alpha ) \alpha ^{2 N} (\gamma  \log (\alpha )+3)+A B \alpha ^N (N \log (\alpha ) (4 \gamma +N (\log (\alpha )+6) \log (N)-2)+6 N-1)+B^2 (2 \gamma +(6 N-1) \log (N)-1)\right)^2}{4 N \left(A \alpha ^N+B \log (N)\right) \left(A N \log (\alpha ) \alpha ^N+B\right) \left(A N \alpha ^N (\gamma  \log (\alpha )+3)+B \gamma +3 B N \log (N)\right)^2} \right]  \\+  2\left[ \frac{\splitfrac{ \splitfrac{ A^3 B N \log (\alpha ) \alpha ^{3 N}}{ \left( \splitfrac{ \splitfrac{ N \log (\alpha ) (N \log (\alpha ) (\gamma  (5.\, -8. \gamma )}{+N \log (N) ((0.5\, -2.5 \gamma )}}{\splitfrac{ \log (\alpha )-6. \gamma -1.5)-1.5) +N (3.\, -18. \gamma )}{+2.5 \gamma -1.5)+1.5 N-1.}}\right) }}{ \splitfrac{ +A^2 B^2 \alpha ^{2 N}}{ \left(N \log (\alpha )  \left(\splitfrac{\splitfrac{ N (\gamma  (12.5\, -12. \gamma )-3.5) \log (\alpha )}{+\log (N) (N \log (\alpha ) (N (1.\, -5. \gamma ) \log (\alpha )}}{\splitfrac{ \splitfrac{ +2.5 \gamma -18. \gamma  N+1.5 N-2.)+3. N-2.)}{+5. \gamma +N^3 (-0.25 \log (\alpha )-1.5)}}{ \log ^2(\alpha ) \log ^2(N)-18. \gamma  N+4.5 N-2.}}\right)  +1.5 N-0.75\right)}}\splitfrac{ +A B^3 \alpha ^N }{\left(\splitfrac{\splitfrac { \splitfrac { \log (N) (N \log (\alpha ) (N (1.\, -2.5 \gamma )}{\times \log (\alpha )+5. \gamma -18. \gamma  N}}{\splitfrac{ +4.5 N-2.)+3. N-1.5)+N (\gamma }{ (10.\, -8. \gamma )-3.) \log (\alpha )}}}{\splitfrac{+2.5 \gamma +N \log (\alpha ) \log ^2(N) (N \log (\alpha ) (-0.5 N \log (\alpha )-1.5 N-0.5)}{+1.5 N-1.)-6. \gamma  N+1.5 N-1.}} \right)}+\splitfrac{\splitfrac{ B^4 (\gamma  (2.5\, -2. \gamma )+\log (N) }{\times(N (1.5\, -6. \gamma )+2.5 \gamma }}{\splitfrac{+(1.5 N-0.75) \log (N)-1.)-0.75)+A^4 \gamma  N^4}{ \log ^3(\alpha ) \alpha ^{4 N} (-2. \gamma  \log (\alpha )-6.)}}}{\splitfrac{ N \left(A \alpha ^N+B \log (N)\right) \left(A N \log (\alpha ) \alpha ^N+B\right)^2}{\times \left(A \gamma  N \log (\alpha ) \alpha ^N+3. A N \alpha ^N+B \gamma +3. B N \log (N)\right)}}\right].
\end{gather}
Similar to the first model (\ref{1st}), we shall now show the viability of the model (\ref{2nd}) in yielding viable estimates of the inflation related observables, for which we shall assume the following combination of the free parameters, 
\begin{equation}
\alpha=1.00417, \hspace*{0.25in} \gamma=1,  \hspace*{0.25in} A= -5.9, \hspace*{0.25in} \text{and} \hspace*{0.25in} B=1,
\end{equation}
which produce the following values of the inflation observables, 
\begin{equation}
r=0.0688202, \hspace*{0.25in} \text{and} \hspace*{0.25in} n_{S}=0.968517.
\end{equation}
Akin to the first model, the theoretical predictions of the inflation observables are in excellent agreement with observations. The expressions for the mimetic potential and the Lagrange multiplier for this model read respectively as,
\begin{equation}
V = 6 \left(2 \left(A \alpha ^N+B \log (N)\right)^{2 \gamma }+\gamma  \left(A \log (\alpha ) \alpha ^N+\frac{B}{N}\right) \left(A \alpha ^N+B \log (N)\right)^{2 \gamma -1}\right)\\ \notag -2 \left(3 \left(A \alpha ^N+B \log (N)\right)^{2 \gamma }+\gamma  \left(A \log (\alpha ) \alpha ^N+\frac{B}{N}\right) \left(A \alpha ^N+B \log (N)\right)^{2 \gamma -1}\right)+4 \rho  \chi -\rho  (4 \chi +1),
\end{equation}
and
\begin{equation}
\lambda = -0.5 \left(6 \left(2 \left(A \alpha ^N+B \log (N)\right)^{2 \gamma }+\gamma  \left(A \log (\alpha ) \alpha ^N+\frac{B}{N}\right) \left(A \alpha ^N+B \log (N)\right)^{2 \gamma -1}\right)+4 \rho  \chi \right)\\ \notag +0.5 \left(6 \left(2 \left(A \alpha ^N+B \log (N)\right)^{2 \gamma }+\gamma  \left(A \log (\alpha ) \alpha ^N+\frac{B}{N}\right) \left(A \alpha ^N+B \log (N)\right)^{2 \gamma -1}\right)-2 \left(3 \left(A \alpha ^N+B \log (N)\right)^{2 \gamma }+\gamma  \left(A \log (\alpha ) \alpha ^N+\frac{B}{N}\right) \left(A \alpha ^N+B \log (N)\right)^{2 \gamma -1}\right)+4 \rho  \chi -\rho  (4 \chi +1)\right)\\ \notag +3 \left(\left(A \alpha ^N+B \log (N)\right)^{2 \gamma }+\gamma  \left(A \log (\alpha ) \alpha ^N+\frac{B}{N}\right) \left(A \alpha ^N+B \log (N)\right)^{2 \gamma -1}\right)+\rho  \chi +\rho  (\chi +0.5).
\end{equation}
\subsection{Inflationary Model III}
We shall now consider a third viable inflationary model which can be expressed as
\begin{equation}\label{3rd}
H(N)=\left(A e^{\beta  N}+B \log N\right)^{\gamma },
\end{equation}
where $A$, $B$ and $\gamma$ are free parameters. Substituting Eq. \ref{3rd} in \ref{slow1} and \ref{slow2}, the slow-roll parameters for the model read
\begin{equation}
\epsilon=-\frac{\gamma  \left(2 A^2 \beta  N^2 (\beta  \gamma +3) e^{2 \beta  N}+A B e^{\beta  N} \left(\beta  (\beta +6) N^2 \log (N)+N (4 \beta  \gamma -2 \beta +6)-1\right)+B^2 (2 \gamma +(6 N-1) \log (N)-1)\right)^2}{4 N \left(A \beta  N e^{\beta  N}+B\right) \left(A e^{\beta  N}+B \log (N)\right) \left(A N (\beta  \gamma +3) e^{\beta  N}+B \gamma +3 B N \log (N)\right)^2},
\end{equation}
and
\begin{equation}
\eta = -\frac{6. \left(B \log (N) \left(B^3 \splitfrac{\splitfrac{\splitfrac{( (0.125\, -0.25 N) \log (N)-0.42 \gamma}{ +(1. \gamma -0.25) N+0.17 )}}{+\splitfrac{ \splitfrac{ A^3 \beta ^3 N^4  (\beta  (0.42 \gamma -0.082)+1. \gamma +0.25)  e^{3 \beta  N}}{+0.042 A^2 \beta  B N e^{2 \beta  N} }}{  \splitfrac{( \beta ^2 (1. \beta +6.) N^3 \log (N)+N (\beta  (-10. \gamma }{+N (20. \beta  \gamma -4. \beta +72. \gamma -6.)+8.)-12.)+8. )}}}}{ \splitfrac{ \splitfrac{ +0.083 A B^2 e^{\beta  N} (N (\beta  (-10. \gamma +N (5. \beta }{ \gamma -2. \beta +36. \gamma -9.)+4.)-6.)}}  {\splitfrac {+\beta  N (N (\beta  ((1. \beta +3.)}{ N+1.)-3.)+2.) \log (N)+3.) } }}\right)+\splitfrac{\splitfrac{\splitfrac { A^4 \beta ^3 \gamma  }{N^4 (0.3 \beta  \gamma +1.)  }}{\splitfrac{  e^{4 \beta  N}}{+A^3 \beta  B N e^{3 \beta  N}}} \splitfrac{ \splitfrac{(N (\beta  (-0.42 \gamma +N (\beta }{(\gamma  (1.3 \gamma -0.8)+0.25)+3. \gamma -0.5)+0.25)-0.25)+0.17)}}{\splitfrac{+A^2 B^2 e^{2 \beta  N} (N (\beta }{ \splitfrac{(-0.8 \gamma +N (\beta  (\gamma  (2. \gamma -2.08)}{+0.59)+3. \gamma -0.75)+0.34)-0.25)+0.125)}}}}{\splitfrac{ +A B^3 e^{\beta  N} (-0.42 \gamma +N (\beta  (\gamma  }{(1.3 \gamma -1.7)+0.5)+1. \gamma -0.25)+0.17)}\splitfrac{+B^4 ((0.3 \gamma}{ -0.42) \gamma +0.125)}    } \right)}   {N \left(A \beta  N e^{\beta  N}+B\right)^2 \left(A e^{\beta  N}+B \log (N)\right) \left(A N (\beta  \gamma +3.) e^{\beta  N}+B \gamma +3. B N \log (N)\right)}.
\end{equation}
Finally, the expressions of the inflationary observables (Eq. \ref{inf}) read
\begin{gather}\label{3r}
r= 16 \left[-\frac{\gamma  \left(2 A^2 \beta  N^2 (\beta  \gamma +3) e^{2 \beta  N}+A B e^{\beta  N} \left(\beta  (\beta +6) N^2 \log (N)+N (4 \beta  \gamma -2 \beta +6)-1\right)+B^2 (2 \gamma +(6 N-1) \log (N)-1)\right)^2}{4 N \left(A \beta  N e^{\beta  N}+B\right) \left(A e^{\beta  N}+B \log (N)\right) \left(A N (\beta  \gamma +3) e^{\beta  N}+B \gamma +3 B N \log (N)\right)^2} \right],
\end{gather}
and
\begin{gather}\label{3ns}
n_{s}= \notag 1- 6\left[ -\frac{\gamma  \left(2 A^2 \beta  N^2 (\beta  \gamma +3) e^{2 \beta  N}+A B e^{\beta  N} \left(\beta  (\beta +6) N^2 \log (N)+N (4 \beta  \gamma -2 \beta +6)-1\right)+B^2 (2 \gamma +(6 N-1) \log (N)-1)\right)^2}{4 N \left(A \beta  N e^{\beta  N}+B\right) \left(A e^{\beta  N}+B \log (N)\right) \left(A N (\beta  \gamma +3) e^{\beta  N}+B \gamma +3 B N \log (N)\right)^2}\right]  \\+ 2 \left[-\frac{6. \left(B \log (N) \left(B^3 \splitfrac{\splitfrac{\splitfrac{( (0.125\, -0.25 N) \log (N)-0.42 \gamma}{ +(1. \gamma -0.25) N+0.17 )}}{+\splitfrac{ \splitfrac{ A^3 \beta ^3 N^4  (\beta  (0.42 \gamma -0.082)+1. \gamma +0.25)  e^{3 \beta  N}}{+0.042 A^2 \beta  B N e^{2 \beta  N} }}{  \splitfrac{( \beta ^2 (1. \beta +6.) N^3 \log (N)+N (\beta  (-10. \gamma }{+N (20. \beta  \gamma -4. \beta +72. \gamma -6.)+8.)-12.)+8. )}}}}{ \splitfrac{ \splitfrac{ +0.083 A B^2 e^{\beta  N} (N (\beta  (-10. \gamma +N (5. \beta }{ \gamma -2. \beta +36. \gamma -9.)+4.)-6.)}}  {\splitfrac {+\beta  N (N (\beta  ((1. \beta +3.)}{ S+1.)-3.)+2.) \log (S)+3.) } }}\right)+\splitfrac{\splitfrac{\splitfrac { A^4 \beta ^3 \gamma  }{S^4 (0.3 \beta  \gamma +1.)  }}{\splitfrac{  e^{4 \beta  S}}{+A^3 \beta  B S e^{3 \beta  S}}} \splitfrac{ \splitfrac{(S (\beta  (-0.42 \gamma +S (\beta }{(\gamma  (1.3 \gamma -0.8)+0.25)+3. \gamma -0.5)+0.25)-0.25)+0.17)}}{\splitfrac{+A^2 B^2 e^{2 \beta  S} (S (\beta }{ \splitfrac{(-0.8 \gamma +S (\beta  (\gamma  (2. \gamma -2.08)}{+0.59)+3. \gamma -0.75)+0.34)-0.25)+0.125)}}}}{\splitfrac{ +A B^3 e^{\beta  S} (-0.42 \gamma +S (\beta  (\gamma  }{(1.3 \gamma -1.7)+0.5)+1. \gamma -0.25)+0.17)}\splitfrac{+B^4 ((0.3 \gamma}{ -0.42) \gamma +0.125)}    } \right)}   {S \left(A \beta  S e^{\beta  S}+B\right)^2 \left(A e^{\beta  S}+B \log (S)\right) \left(A S (\beta  \gamma +3.) e^{\beta  S}+B \gamma +3. B S \log (S)\right)} \right].
\end{gather}
With the same line of reasoning, we shall now try to find suitable values of the free parameters to generate observationally acceptable values of the inflation observables. We therefore use the following union of the free parameters as, 
\begin{equation}
\beta=-0.126 \hspace*{0.25in} \gamma=0.03,  \hspace*{0.25in} A= 5500, \hspace*{0.25in} \text{and} \hspace*{0.25in} B=60,
\end{equation}
which give, 
\begin{equation}
r=0.00125178, \hspace*{0.25in} \text{and} \hspace*{0.25in} n_{S}=0.960436.
\end{equation}
The theoretical estimate matches with observations excellently and therefore confirms the viability of the model (\ref{3rd}). Next, the expressions for the mimetic potential and the Lagrange multiplier for the model read respectively as,
\begin{equation}
V = 6 \left(2 \left(A e^{\beta  N}+B \log (N)\right)^{2 \gamma }+\gamma  \left(A \beta  e^{\beta  N}+\frac{B}{N}\right) \left(A e^{\beta  N}+B \log (N)\right)^{2 \gamma -1}\right)\\ \notag -2 \left(3 \left(A e^{\beta  N}+B \log (N)\right)^{2 \gamma }+\gamma  \left(A \beta  e^{\beta  N}+\frac{B}{N}\right) \left(A e^{\beta  N}+B \log (N)\right)^{2 \gamma -1}\right)+4 \rho  \chi -\rho  (4 \chi +1),
\end{equation}
and
\begin{equation}
\lambda = -0.5 \left(6 \left(2 \left(A e^{\beta  N}+B \log (N)\right)^{2 \gamma }+\gamma  \left(A \beta  e^{\beta  N}+\frac{B}{N}\right) \left(A e^{\beta  N}+B \log (N)\right)^{2 \gamma -1}\right) +4 \rho  \chi \right)\\ \notag+0.5 \left(6 \left(2 \left(A e^{\beta  N}+B \log (N)\right)^{2 \gamma }+\gamma  \left(A \beta  e^{\beta  N}+\frac{B}{N}\right) \left(A e^{\beta  N}+B \log (N)\right)^{2 \gamma -1}\right)-2 \left(3 \left(A e^{\beta  N}+B \log (N)\right)^{2 \gamma }+\gamma  \left(A \beta  e^{\beta  N}+\frac{B}{N}\right) \left(A e^{\beta  N}+B \log (N)\right)^{2 \gamma -1}\right)+4 \rho  \chi -\rho  (4 \chi +1)\right)\\ \notag+3 \left(\left(A e^{\beta  N}+B \log (N)\right)^{2 \gamma }+\gamma  \left(A \beta  e^{\beta  N}+\frac{B}{N}\right) \left(A e^{\beta  N}+B \log (N)\right)^{2 \gamma -1}\right)+\rho  \chi +\rho  (\chi +0.5).
\end{equation}

\section{Conclusions}\label{sec4}
Inflation provides a clear and coherent description of the formation of large-scale structures, CMB anisotropies, and provides solutions to the flatness, and horizon problems and the fine-tuning problem \cite{in1,in2,in3}. In Einstein's gravity, a scalar field called "inflaton" is predicted to be responsible for carrying out an inflation \cite{in1}. \\
Modified theories of gravity have been very productive in delineating cosmological problems such as the flat rotation curves of spiral galaxies or the late-time cosmic acceleration without requiring dark matter and dark energy \cite{bp2,bp3}. The gravitational interactions in extended theories of gravity are scale-dependent but conserve the results of GR at the Solar System scales \cite{bp2}. Thus, Einstein's gravity can be considered as being a specific case of a more diverse set of gravitational theories \cite{bp}. \\
Mimetic gravity is a well-organized Weyl-symmetric augmentation of Einstein's gravity connected through a singular disformal transformation capable of presenting a unified description of the evolution of the Universe from the primordial inflation to the late-time acceleration \cite{mimetic}. In this paper, we have shown that by employing mimetic $f(R,T)$ gravity coupled with Lagrange multiplier and mimetic potential, viable inflationary cosmological solutions consistent with the latest Planck and BICEP2/Keck Array data are possible. We present here three viable inflationary solutions represented by $H(N)=\left(A \exp \beta  N+B \alpha ^N\right)^{\gamma }$, $H(N)=\left(A \alpha ^N+B \log N\right)^{\gamma }$, and $H(N)=\left(A e^{\beta  N}+B \log N\right)^{\gamma }$, where $A$, $\beta$, $B$, $\alpha$ and $\gamma$ are free parameters. We carried out the analysis with the simplest minimal $f(R,T)$ function of the form $f(R,T)= R + \chi T$, where $\chi$ is the model parameter. We report that for the chosen $f(R,T)$ gravity model, viable cosmologies are obtained compatible with observations by conveniently setting the Lagrange multiplier and the mimetic potential.

\section*{Acknowledgments}

We thank the anonymous reviewers and the editor for their helpful suggestions.

\end{document}